\documentclass{article}
\usepackage{spconf,amsmath,graphicx}
\usepackage{amssymb}
\usepackage{bbm}
\usepackage{multirow}
\usepackage{booktabs}
\usepackage[inline]{enumitem}

\title{A Framework for Unified Real-time Personalized and Non-Personalized Speech Enhancement}
\name{Zhepei Wang$^{\sharp}$ \sthanks{Work performed while at Amazon Web Services.}, Ritwik Giri$^{\dagger}$, Devansh Shah$^{\dagger}$, Jean-Marc Valin$^{\dagger}$, Michael M. Goodwin$^{\dagger}$, Paris Smaragdis$^{\sharp \dagger}$ }
\address{$^{\dagger}$Amazon Web Services\\
$^{\sharp}$University of Illinois at Urbana-Champaign
}
\begin{document}
\ninept
\maketitle
\begin{abstract}
In this study, we present an approach to train a single speech enhancement network that can perform both personalized and non-personalized speech enhancement. This is achieved by incorporating a frame-wise conditioning input that specifies the type of enhancement output. To improve the quality of the enhanced output and mitigate oversuppression, we experiment with re-weighting frames by the presence or absence of speech activity and applying augmentations to speaker embeddings. By training under a multi-task learning setting, we empirically show that the proposed unified model obtains promising results on both personalized and non-personalized speech enhancement benchmarks and reaches similar performance to models that are trained specialized for either task. The strong performance of the proposed method demonstrates that the unified model is a more economical alternative compared to keeping separate task-specific models during inference.

\end{abstract}
\newcommand{\R}{\mathbb{R}}
\newcommand{\mone}{\mathbbm{1}}
\newcommand{\bzero}{\mathbf{0}}
\newcommand{\bD}{\mathbf{D}}
\newcommand{\bH}{\mathbf{H}}
\newcommand{\bM}{\mathbf{M}}
\newcommand{\bS}{\mathbf{S}}
\newcommand{\bU}{\mathbf{U}}
\newcommand{\bV}{\mathbf{V}}
\newcommand{\bW}{\mathbf{W}}
\newcommand{\bZ}{\mathbf{Z}}
\newcommand{\ba}{\mathbf{a}}
\newcommand{\bh}{\mathbf{h}}
\newcommand{\bq}{\mathbf{q}}
\newcommand{\bs}{\mathbf{s}}
\newcommand{\bv}{\mathbf{v}}
\newcommand{\bx}{\mathbf{x}}
\newcommand{\by}{\mathbf{y}}
\newcommand{\bz}{\mathbf{z}}
\newcommand{\calE}{\mathcal{E}}
\newcommand{\calL}{\mathcal{L}}
\newcommand{\cat}{\text{Concat}}

\section{Introduction}
\label{sec:intro}

Online teleconference systems have become a preferred way of communication in recent years when in-person meetings or conferences are costly or infeasible. However, these remote conversations often take place in a noisy environment, and it is challenging to preserve the intelligibility of speech in the presence of ambient noise. To improve the communication experience, speech enhancement techniques have become pivotal in online meeting systems to filter out background noise and improve speech quality. While advanced deep neural network architectures have achieved state-of-the-art in offline speech enhancement tasks \cite{dc_unet, poconet}, recent advances in speech enhancement have been focused on efficient model designs that enable high-quality noise reduction in real-time \cite{percepnet, remixit, fullsubnet, fullsubnet+, frcrn}.

Despite their ability to clean up environmental noise, these systems are not effective to remove human voices from the background. Consider a customer service representative who answers phone calls in an open office setting surrounded by a group of other talkers. Personalized speech enhancement will be applicable to this scenario by extracting the voice of this representative while suppressing all other speakers and the background noise. Deep-learning-based personalized speech enhancement systems usually involve, 
\begin{enumerate*}[label=(\roman*)]
    \item a speaker embedder module that provides cues for the speaker in interest;
    \item a speech enhancement module to recover the target speech from the input mixture with ambiance noise and interference.
\end{enumerate*} These modules can be learned either separately \cite{VF, speakerbeam, TencentTSE, exformer} or jointly with each other \cite{td_speakerbeam, time_tse_mult, speakerfilter, spex, spex+, spex++}. With consideration of run-time complexity and memory usage, a number of real-time personalized speech enhancement methods have shown promising results in efficiently extracting the target speaker's voice \cite{VF_lite, pers_percepnet, tea-pse, multistage_pse_kuaishou}.

While personalized speech enhancement systems provide an improved conversation experience for the target user, non-personalized speech enhancement systems are still required in a variety of use cases. For instance, personalized systems are not applicable when the information of the target speaker is unknown in advance. There are also scenarios in which we need to preserve speech from all talkers in the same meeting room. Even more challenging are the time-varying circumstances in which we want to first perform non-personalized enhancement to allow the speech of multiple speakers through, and switch later to focus on a particular speaker with personalized enhancement. Conventionally, to enable both personalized and non-personalized speech enhancement, we train a separate model for each application. During inference, it typically requires us to maintain both models in memory and choose the appropriate network depending on the use cases. With the majority of model components being identical between personalized and non-personalized enhancement networks, it naturally brings in the question of whether it is feasible to obtain a single enhancement model for both tasks.

\begin{figure}[!tb]
    \centering
    \includegraphics[width=0.48\textwidth]{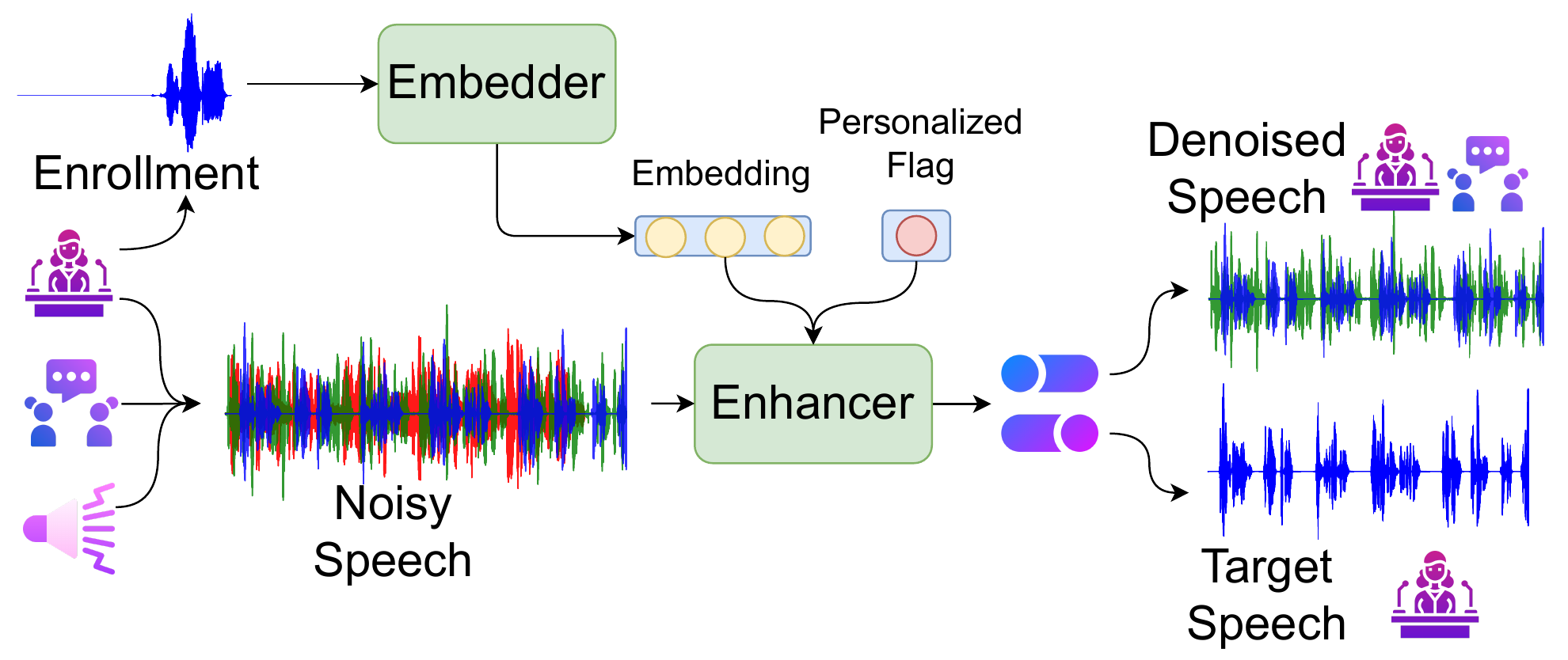}
    \caption{The pipeline of the Unified PercepNet. The enhancer is conditioned by a binary input to determine whether personalized or non-personalized output is produced.}
    \label{fig:pipeline}
\end{figure}

In this work, we present the Unified PercepNet (UPN), illustrated in Fig. \ref{fig:pipeline}, which is trained with multi-task learning to achieve both personalized and non-personalized speech enhancement. The UPN consists of an embedder and an enhancer network. The enhancer is controlled by a user-provided input that specifies the output behavior. When the personalized mode is activated, the model is expected to attend to the target speaker and clean up both background noise and any interference speech; otherwise, it should behave as a non-personalized enhancement model that suppresses only the environmental sound. All parameters of the enhancement network are shared between both tasks. While the network proposed in \cite{multistage_pse_kuaishou} can also perform both personalized and non-personalized enhancement, our proposed method is distinct in the perspectives of,
\begin{enumerate*}[label=(\roman*)]
    \item we train the model with personalized and non-personalized data jointly instead of in separate stages;
    \item we provide frame-wise personalized or non-personalized control to enable switching of the enhancement mode within a single input sequence.
\end{enumerate*}

Additionally, we examine the effectiveness of several approaches to improve the quality of enhancement. We consider using a loss function with weight adjustment between voiced and unvoiced frames in the target and applying data augmentation to the speaker embeddings to combat oversuppression and overfitting. We evaluate the model's performance on both personalized and non-personalized tracks from the 4th Deep Noise Suppression Challenge \cite{dns_4} using an identical model for both tasks. The overall quality of the enhanced output of the UPN is comparable to either the personalized or non-personalized model trained under a single-task setting with the specialized dataset for the corresponding task. 
To the best of our knowledge, this is the first study that considers training a unified model for both personalized and non-personalized speech enhancement under a joint multi-task learning setting.

\section{Proposed Framework}
\label{sec:method}

\subsection{Speaker Embedding Network}
\label{ssec:method_embedder}
The speaker embedding network is used for extracting cues for the target speaker from an enrollment utterance.
This process can be formulated as $\bz = \calE(\bx_e)$, where $\bx_e$ is the enrollment signal, $\bz \in \R^D$ is a $D$-dimensional vector representing the speaker's identity, and $\calE$ is the embedding network. The embedding $\bz$ is used as a conditioning input to the enhancement model.

We choose the ECAPA-TDNN \cite{ecapa_tdnn} architecture for the speaker embedder as it achieves state-of-the-art results on several speaker recognition tasks \cite{voxceleb, voxceleb2, voxsrc_19}. The embedder network is trained with the AAM-softmax loss \cite{arcface_cv, aam_softmax} before the training of the enhancer.
The weights of this pre-trained embedder remain unchanged during the training of the enhancement model.



\subsection{Enhancement Network}
\label{ssec:method_enhancer}



\subsubsection{Model Architecture}
\label{sssec:method_upn_arch}
The architecture of the enhancement network of UPN is based on the PercepNet \cite{percepnet} and the Personalized PercepNet (PPN) \cite{pers_percepnet}.
The input features are obtained from 32 bands following the equivalent rectangular bandwidth (ERB) scale, where for each band we compute two features: the magnitude and the pitch coherence values. Along with four additional general features including the pitch period, we obtain a 68-dimensional input for the model. The model consists of two 1D convolutional layers followed by a few blocks of Gated recurrent units (GRUs). For each band $b$, it computes the following quantities for each frame $t$: 
\begin{enumerate*}[label=(\roman*)]
    \item $\hat{g}_{b, t}$, the gain of the estimated enhanced speech in the form of the ratio mask to be applied to the magnitude of the input mixture;
    \item $\hat{r}_{b, t}$, the pitch-filter strengths, which is used to control the strength of a comb filter applied to the time-domain reconstruction of the output signal to further reduce noise between pitch harmonics.
\end{enumerate*}

The PPN is the personalized variant of the PercepNet. It takes the speaker embedding $\bz$ as a conditioning input, which is concatenated to the mixture representation at the beginning of the GRU layer.
It also computes the frame-wise voice activity detection (VAD) score $\hat{y}_t \in [0, 1]$ as the probability of the target speaker present in the enhanced output at frame $t$. 

\begin{figure}[!tb]
     \centering
     \includegraphics[width=0.48\textwidth]{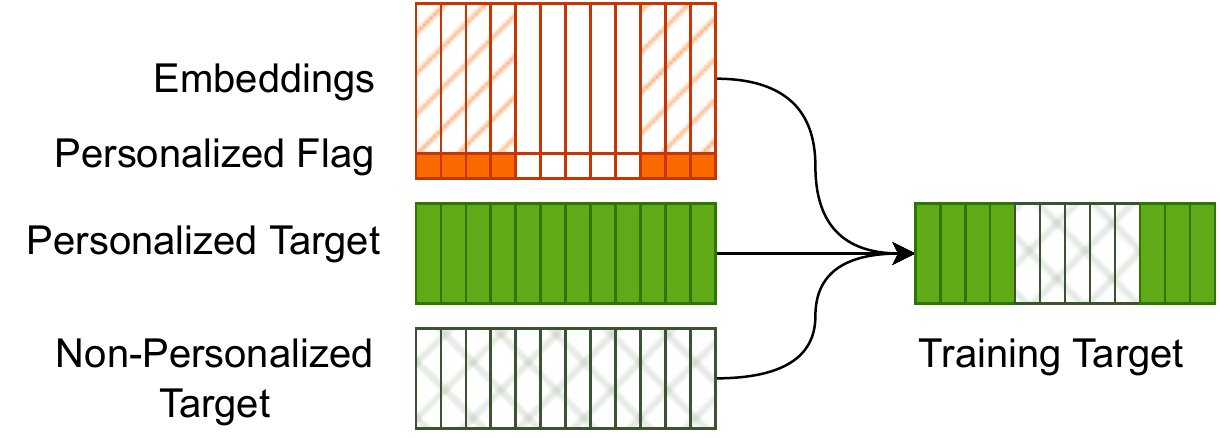}
     \caption{Training targets of the UPN. For the frames where the personalized flag is activated, the personalized reference speech is used; otherwise, the non-personalized target is selected.}
     \label{fig:dataloader}
 \end{figure}

The UPN further extends the PPN by enabling both personalized and non-personalized speech enhancement in a unified model. This is achieved by incorporating a personalized flag $q \in \{0, 1\}$ into the enhancement network, which is a single binary bit to control the behavior of speech enhancement. The personalized flag is first appended to the speaker embedding $\bz$ to form the personalized-controlled embedding $\bz' \in \R^{D+1}$. If this flag is activated (i.e., $q=1$), the enhancement model should extract speech from only the target speaker. Otherwise, the model should perform non-personalized speech enhancement by filtering out the background noise only.
For mixtures with multiple speakers, the network does not distinguish between primary and interference speakers and instead retains all the speech signals. Following the PPN, we concatenate $\bz'$ to each frame of the latent representations of the mixture; however, as opposed to the speaker embedding $\bz$ being invariant over time, it is possible to switch the personalized control $q$ between frames. At each frame $t$, the concatenation of the speaker embedding and personalized flag is hence defined as 
\begin{align}
    \bz'_t = \begin{cases}
     \cat{(\bz, q_t)}, \quad \text{if } q_t = 1, \\
     \cat{(\bzero^D, q_t)} =  \bzero^{D+1} \quad \text{if } q_t = 0,
    \end{cases}
\label{eq:emb_pflag_cat}
\end{align}
 where $\bzero^D$ is a $D$-dimensional zero vector. We adapt the training data to this frame-level personalized control as shown in Fig. \ref{fig:dataloader}, where we use the personalized reference (speech from only the target speaker) as the training target for frames with $q_t = 1$ and the non-personalized ground-truth (speech from all speakers) otherwise. This provides the flexibility to toggle between personalized and non-personalized enhancement within a single input sequence.
 \vspace{-0.5em}
 \vspace{-0.5em}
 \subsubsection{Loss Functions}
 \label{sssec:method_loss}
 \vspace{-0.5em}
We adpot the loss functions for the gain $\calL_g(b, t)$, pitch-filter strength $\calL_r(b, t)$, and VAD score $\calL_v(t)$ from the training of the PercepNet \cite{percepnet} and the PPN \cite{pers_percepnet}. Both $\calL_g(b,t)$ and $\calL_r(b, t)$ are averaged across all bands and frames, and $\calL_v(t)$ is averaged across frames.

One primary obstacle to the speech enhancement system is the oversuppression problem where the desired speech is also muted along with the noise or interference in the estimated output \cite{tea-pse}. To address this challenge, we modify the loss functions by increasing the weights on the voiced frames using the ground-truth VAD scores $y_t$ as follows:
\begin{align}
\begin{split}
     \calL_G & = \frac{1}{BT}\sum_{b,t} \Bigl\{ \bigl( \mu \mone[y_t=1] + (1-\mu) \mone[y_t=0] \bigr) \cdot \calL_g(b, t) \Bigr\} , \\
     \calL_R & = \frac{1}{BT}\sum_{b,t} \Bigl\{ \bigl(\mu \mone[y_t=1] + (1-\mu)\mone[y_t=0] \bigr) \cdot \calL_r(b, t) \Bigr\}, \\
     \calL_V & = \frac{1}{T}\sum_{t} \Bigl\{ \bigl(\mu \mone[y_t=1] + (1-\mu )\mone[y_t=0] \bigr) \cdot \calL_v(t) \Bigr\},
\end{split}
   \label{eq:loss_vad_weight}
\end{align}
where $\mu \in [0, 1]$ is the hyperparameter to control the importance of frames with and without (target) speech activity. The VAD-weighted overall objective is expressed as $\calL_\text{VAD-weighted} = \calL_G + \calL_R + \calL_V$, and it is equivalent to the original loss at $\mu=0.5$. As $\mu$ increases from 0.5 to 1, the model is encouraged to attend more to the prediction of frames where the enhanced speech should not be silent as a way to combat oversuppression.
\section{Experimental Configurations}
\label{sec:exp}
\vspace{-0.5em}

\subsection{Training of the Speaker Embedder}
\label{ssec:exp_emb}

We pre-train the speaker embedder network with the VoxCeleb1 \cite{voxceleb}, VoxCeleb2 \cite{voxceleb2}, and LibriSpeech \cite{librispeech} datasets. We follow \cite{ecapa_tdnn} to configure the architecture of the ECAPA-TDNN using three SE-Res2Block with 1024 channels and an output dimension $D=192$. This pre-trained embedder model obtains an equal error rate (EER) of 0.39\% when evaluated on a text-independent speaker verification task on the VoxCeleb1 test pairs \cite{voxceleb}.

\subsection{Training of the Enhancement Network}
\label{ssec:exp_enhancement}

\subsubsection{Data Preprocessing}
\label{sssec:exp_enhancement_data}

We construct the training set for the enhancement model based on the VoxCeleb1, VoxCeleb2, and LibriSpeech datasets using the same data pre-processing technique described in \cite{poconet, pers_percepnet}. 
In addition, we also train the enhancement network with the personalized track of the dataset for the 4th Deep Noise Suppression Challenge \cite{dns_4} (PDNS), where the training set contains recordings from 3,230 speakers with a total duration of 750 hours with a sampling rate of 48 kHz. We discard speakers with singing voice only, with less than 300s of training utterances, or with less than 60s of enrollment speech. Combining all datasets, we have more than 7,000 speakers.


\subsubsection{Semi-supervised Data Cleanup: Speaker Labeling Issues}
\label{sssec:exp_enhancement_semisup}

With initial manual inspections of the PDNS dataset, we notice several mismatches between the speaker labels and the actual speaker from the recordings. There are several training or enrollment recordings with speech from multiple speakers, and there are also instances where the same speaker is labeled with more than one speaker id.
To filter out recordings with problematic annotations, we develop a semi-supervised procedure using the pre-trained speaker embedding network. To detect utterances with multiple speakers, we first segment each clip into 30-second overlapping chunks and obtain an embedding for each chunk using the pre-trained embedder. We then compute the cosine similarity scores for each pair of embeddings.
If the average is lower than 0.8, we identify this clip as a multi-speaker utterance and will discard it from training. To detect duplicate clips for the identical speaker, we compute the similarity scores between embeddings obtained from different clips. For a given target speaker, we only select interference speakers whose average similarity score between this target speaker is less than 0.5.


\subsubsection{Mixture Synthesis and Data Augmentation}
\label{sssec:exp_enhancement_aug}
To create noisy input speech, we mix the clean speech signals with the noise data in \cite{percepnet, pers_percepnet} which contains 80 hours of various noise types at a sampling rate of 48 kHz. The signal-to-noise ratio (SNR) of the synthesized mixture is uniformly distributed between -5 dB and 35 dB. For each speaker, we generate 80 mixture segments, each with a duration of 80 seconds, where 40 segments contain overlapping speech with a maximum of two speakers at any given time, 20 segments are the alternating but non-overlapping speech of multiple speakers, and the rest 20 segments contain only a single speaker. For multi-speaker mixtures, the signal-to-interference ratio (SIR) is between -2 dB and 10 dB.
We also apply the following data augmentation techniques including reverberation, random low-pass filtering, random EQ, and random level.
    
We also experiment with data augmentations on the enrollment utterances to mitigate overfitting in the embedding space.
For each enrollment utterance, we generate 10 different variants by applying random noise and reverberation. The pre-trained embedder achieves an EER of 0.614\% on the augmented enrollment speech compared to 0.341\% on the original set. Despite a higher EER, this value still indicates the exceptional ability of the embeddings to discriminate the identities of different speakers.

\subsubsection{Training Setup}
\label{sssec:exp_enhancement_setup}
Before model training, we precompute the features for the input and target from the generated mixture and reference clips using a frame size of 10 ms and a look-ahead window of 30 ms, and the length of the input is equivalent to 20s of audio. All recordings are resampled to 48 kHz before feature computation. The personalized flag $q_t$ is generated on the fly with an equal probability for the following options:
\begin{enumerate*}[label=(\roman*)]
    \item full personalized mode, where $q_t=1$ for all frames $t$;
    \item full non-personalized mode, where $q_t=0$ for $t$;
    \item alternating controls, where $q_t$ changes between 0 and 1 for either one or two switches, and $q_t$ must remain constant for at least 200 frames (2 seconds).
\end{enumerate*}


The architecture of the enhancement model follows the PPN-1024 in \cite{pers_percepnet}. We train the model using a batch size of 256 on 8 NVIDIA-V100 GPUs.

\section{Results and Discussions}
\label{sec:res}
\vspace{-0.5em}
\begin{table}[!htb]
    \centering
    \footnotesize
    \begin{tabular}{c|c|c|c|c|c|c}
    \toprule
     \multirow{2}{*}{Method}   &  \multicolumn{3}{c|}{Personalized} & \multicolumn{3}{c}{Non-personalized} \\
         & SIG & BAK & OVRL & SIG & BAK & OVRL \\
         \midrule
         Noisy Input & \textbf{3.814} & 2.23 & 2.418 & 2.988 & 2.559 & 2.206 \\
         PercepNet & 3.622 & 3.079 & 2.723 & \textbf{3.089} & 3.875 & \textbf{2.752} \\
         PPN & 3.427 & 3.659 & \textbf{2.880} & - & - & - \\
         \midrule
         UPN-OE-0.5 & 3.285 & 3.661 & 2.746 & 3.065 & \textbf{3.923} & 2.745 \\
         UPN-OE-0.75 & 3.358 & 3.600 & 2.787 & 3.071 & 3.902 & 2.744 \\
         UPN-OE-0.9 & 3.346 & 3.621 & 2.796 & 3.075 & 3.899 & {2.748} \\
         UPN-AE-0.5 & 3.33 & \textbf{3.674} & 2.800 & 3.059 & 3.922 & 2.739 \\
         UPN-AE-0.9 & {3.454} & 3.607 & {2.877} & {3.082} & 3.884 & 2.747 \\
         \bottomrule
    \end{tabular}
    \caption{Test results on the personalized and non-personalized test set with DNSMOS P.835. For the unified models, ``OE'' represents training with the original enrollment speech, and ``AE'' is using augmented enrollment speech. The numbers (0.5, 0.75, 0.9) refer to the coefficient $\mu$ for weighting the voiced frames. The task-specific models (PercepNet and PPN) are trained with $\mu=0.9$, and enrollment speech augmentation is applied to the training of PPN.}
    \label{tab:dnsmos}
\end{table}

\begin{table}[!htb]
    \centering
    \begin{tabular}{c|c|c}
    \toprule
     Method & Personalized & Non-personalized \\
         \midrule
         Noisy Input &2.88 &3.18  \\
         PercepNet &- &\textbf{3.43} \\
         PPN &3.10  & - \\
         UPN-AE-0.9 &\textbf{3.11} &3.40 \\
         \bottomrule
    \end{tabular}
    \caption{Subjective evaluation (MOS) of different methods on the personalized and non-personalized test sets.}
    \label{tab:mos}
\end{table}
\vspace{-0.5em}
We evaluate the UPN with both non-personalized and personalized speech enhancement data from the official development test sets from the DNS Challenge \cite{dns_4}. The non-personalized test set contains 930 real-world recordings of noisy English speech. The personalized track contains 1,443 test clips and each primary speaker also comes with an additional 2.5 minutes of enrollment speech.
We set the flag $q_t = 1$ for all frames $t$ for the personalized output and $q_t = 0$ for the non-personalized ones. The complexity of UPN is mostly dictated by the number of parameters in the DNN model and is the same as PPN-1024 \cite{pers_percepnet}. The model has
26.5M parameters. With a 10-ms frame size,  UPN requires 17.2\% of one mobile x86 core
(1.8 GHz Intel i7-8565U CPU) for real-time operation.

For model selection, we evaluate the enhanced output with DNSMOS P.835 \cite{dnsmos_p835}, a non-intrusive approach that predicts subjective rating of the quality of speech (SIG), suppression of background noise (BAK), and overall quality (OVRL) from poor (score=1) to excellent (score=5) using the local evaluation method from the official GitHub \footnote{https://github.com/microsoft/DNS-Challenge/tree/master/DNSMOS}. The results for both personalized and non-personalized test sets are reported in Table \ref{tab:dnsmos}.


\subsection{Impact of VAD-Weighted Objectives}
\label{ssec:res_vad}
We first conduct an ablation study to assess the impact of the VAD-weighted learning objectives \eqref{eq:loss_vad_weight} with the VAD coefficient $\mu=0.5, 0.75$ and 0.9.
Notice that $\mu$ controls the trade-off between suppressing the background noise and preserving desired speech content. We hypothesize that an increased value of $\mu$ is beneficial for resolving oversuppression since the model learns to attend more to reproducing the speech signal rather than removing noise in the silent frames. For the personalized evaluation (without enrollment augmentation), we notice that 
\begin{enumerate*}[label=(\roman*)]
    \item the signal quality reaches its peak as $\mu$ increases from 0.5 to 0.75 but slightly decreases when $\mu=0.9$;
    \item the reverse trend holds for the background quality;
    \item the best overall quality is obtained at $\mu=0.9$.
\end{enumerate*}
Along with our preliminary listening tests where we observe the greatest amount of oversuppression failure cases with $\mu=0.5$, the experimental results are consistent with our hypothesis. In general, the model's output is more likely to contain leakage of background noise if trained with a larger weight $\mu$, but the drawbacks are compensated by the alleviation of oversuppression, and the overall quality improves.

\subsection{Impact of Enrollment Speech Augmentation}
\label{ssec:res_enrol_aug}
We next study the effects of applying data augmentation to the enrollment utterances before obtaining the speaker embeddings when training with the personalized data. In our initial experiments using the original enrollment speech to obtain embeddings, we detect that the signal quality severely degrades in later epochs. Moreover, when using the same single-speaker mixture as input, the non-personalized output can reconstruct the speech whereas the personalized output suffers from noticeable oversuppression. We suspect that the enhancement model may overfit the embedding space and therefore fail to identify the test speaker under personalized control, and as an attempt to resolve this issue, we augment the enrollment speech with noise and reverberation. Using the personalized benchmark, with $\mu=0.5$ we observe a comprehensive improvement of speech, background, and overall quality after applying enrollment augmentation by 1.37\%, 0.36\%, and 1.97\%, respectively; with $\mu=0.9$, the signal and overall quality increases by 3.23\% and 2.90\% with a slight decrease of 0.39\% in background quality. These results indicate the effectiveness of the augmentation in the embedding space.
As the set of embeddings becomes more diverse, the enhancement model is less prone to overfitting and more robust to potentially test-time speaker embeddings unseen during training.

\subsection{Comparison with Single-Task Models}
\label{ssec:res_single_task}
\vspace{-0.5em}
Lastly, we compare the performance of the proposed unified enhancement model with the task-specific models trained for either personalized or non-personalized speech enhancement. For a fair comparison, these two reference models are trained with a VAD weight $\mu=0.9$, and we apply the same enrollment augmentation to the reference personalized model, PPN. The corresponding unified model, UPN-AE-0.9, is comparable to the PPN under the personalized test cases.
The overall quality scores are nearly identical, and the unified model obtains a signal quality score 0.79\% higher than the PPN. For the non-personalized benchmark, the unified model obtains close figures across all three metrics compared to the reference non-personalized model (0.23\% lower in SIG, 0.23\% higher in BAK, and 0.18\% lower in OVRL). We can see that the proposed UPN trained under the multi-task setting achieves roughly equivalent performance to both personalized and non-personalized models trained with a single task.

We further verify the effectiveness of UPN with Mean Opinion Score (MOS), following the ITU-T P.808 crowdsourcing approach, on the overall quality of speech. We list the results of the mixture input, PercepNet, PPN, and the UPN with $\mu=0.9$ and embedding augmentation in Table \ref{tab:mos}. The observations are similar to the DNSMOS P.835 results, where the UPN obtains a close score to the PercepNet for the non-personalized evaluation while slightly exceeding PPN on the personalized data (all within the 95\% confidence interval of 0.04). The close performance implies the potential to replace the two task-specific enhancement networks with the proposed unified model as a more memory-efficient alternative.
\section{Conclusion}
\label{sec:conclusion}
\vspace{-0.5em}

We propose a framework to train a single model for both personalized and non-personalized speech enhancement tasks under a multi-task learning setting. The types of enhancement output are controlled by a frame-wise conditioning input, and the model is flexible to switch between personalized and non-personalized output within a single input sequence. With the proposed VAD-weighted learning objective and embedding augmentation, the proposed unified model reaches a similar performance to the reference models that are trained with a specialized task either on personalized or non-personalized enhancement. Instead of keeping separate models for each task, our work shows the potential to replace them with a unified model, which reduces the memory for storing models and the resources required to retrain or update the model in the future.

\bibliographystyle{IEEEbib}
\bibliography{refs}

\begin{thebibliography}{10}

\bibitem{dc_unet}
Hyeong-Seok Choi, Jang-Hyun Kim, Jaesung Huh, Adrian Kim, Jung-Woo Ha, and
  Kyogu Lee,
\newblock ``Phase-aware speech enhancement with deep complex u-net,''
\newblock in {\em ICLR}, 2018.

\bibitem{poconet}
Umut Isik, Ritwik Giri, Neerad Phansalkar, Jean-Marc Valin, Karim Helwani, and
  Arvindh Krishnaswamy,
\newblock ``Poconet: Better speech enhancement with frequency-positional
  embeddings, semi-supervised conversational data, and biased loss,''
\newblock in {\em INTERSPEECH}, 2020.

\bibitem{percepnet}
Jean-Marc Valin, Umut Isik, Neerad Phansalkar, Ritwik Giri, Karim Helwani, and
  Arvindh Krishnaswamy,
\newblock ``A perceptually-motivated approach for low-complexity, real-time
  enhancement of fullband speech,''
\newblock in {\em INTERSPEECH}, 2020.

\bibitem{remixit}
Efthymios Tzinis, Yossi Adi, Vamsi~Krishna Ithapu, Buye Xu, Paris Smaragdis,
  and Anurag Kumar,
\newblock ``Remixit: Continual self-training of speech enhancement models via
  bootstrapped remixing,''
\newblock in {\em IEEE ICASSP}, 2022.

\bibitem{fullsubnet}
Xiang Hao, Xiangdong Su, Radu Horaud, and Xiaofei Li,
\newblock ``Fullsubnet: A full-band and sub-band fusion model for real-time
  single-channel speech enhancement,''
\newblock {\em IEEE ICASSP}, 2021.

\bibitem{fullsubnet+}
Jun Chen, Zilin Wang, Deyi Tuo, Zhiyong Wu, Shiyin Kang, and Helen Meng,
\newblock ``Fullsubnet+: Channel attention fullsubnet with complex spectrograms
  for speech enhancement,''
\newblock in {\em IEEE ICASSP}, 2022.

\bibitem{frcrn}
Shengkui Zhao, Bin Ma, Karn~N. Watcharasupat, and Woon-Seng Gan,
\newblock ``Frcrn: Boosting feature representation using frequency recurrence
  for monaural speech enhancement,''
\newblock in {\em IEEE ICASSP}, 2022.

\bibitem{VF}
Quan Wang, Hannah Muckenhirn, Kevin Wilson, Prashant Sridhar, Zelin Wu, John~R.
  Hershey, Rif~A. Saurous, Ron~J. Weiss, Ye~Jia, and Ignacio~Lopez Moreno,
\newblock ``{VoiceFilter: Targeted Voice Separation by Speaker-Conditioned
  Spectrogram Masking},''
\newblock in {\em INTERSPEECH}, 2019.

\bibitem{speakerbeam}
Kateřina Žmolíková, Marc Delcroix, Keisuke Kinoshita, Tsubasa Ochiai,
  Tomohiro Nakatani, Lukáš Burget, and Jan Černocký,
\newblock ``Speakerbeam: Speaker aware neural network for target speaker
  extraction in speech mixtures,''
\newblock {\em IEEE JSTSP}, vol. 13, no. 4, 2019.

\bibitem{TencentTSE}
Xuan Ji, Meng Yu, Chunlei Zhang, Dan Su, Tao Yu, Xiaoyu Liu, and Dong Yu,
\newblock ``Speaker-aware target speaker enhancement by jointly learning with
  speaker embedding extraction,''
\newblock in {\em IEEE ICASSP}, 2020.

\bibitem{exformer}
Zhepei Wang, Ritwik Giri, Shrikant Venkataramani, Umut Isik, Jean-Marc Valin,
  Paris Smaragdis, Mike Goodwin, and Arvindh Krishnaswamy,
\newblock ``Semi-supervised time domain target speaker extraction with
  attention,'' 2022.

\bibitem{td_speakerbeam}
Marc Delcroix, Tsubasa Ochiai, Kateřina Žmol{\'i}kov{\'a}, Keisuke Kinoshita,
  Naohiro Tawara, Tomohiro Nakatani, and Shoko Araki,
\newblock ``Improving speaker discrimination of target speech extraction with
  time-domain speakerbeam,''
\newblock {\em IEEE ICASSP}, 2020.

\bibitem{time_tse_mult}
Jianshu Zhao, Shengzhou Gao, and Takahiro Shinozaki,
\newblock ``{Time-Domain Target-Speaker Speech Separation with Waveform-Based
  Speaker Embedding},''
\newblock in {\em INTERSPEECH}, 2020.

\bibitem{speakerfilter}
Shulin He, Hao Li, and Xueliang Zhang,
\newblock ``Speakerfilter: Deep learning-based target speaker extraction using
  anchor speech,''
\newblock in {\em IEEE ICASSP}, 2020.

\bibitem{spex}
Chenglin Xu, Wei Rao, Eng~Siong Chng, and Haizhou Li,
\newblock ``Spex: Multi-scale time domain speaker extraction network,''
\newblock {\em IEEE/ACM TASLP}, vol. 28, 2020.

\bibitem{spex+}
Meng Ge, Chenglin Xu, Longbiao Wang, Chng~Eng Siong, Jianwu Dang, and Haizhou
  Li,
\newblock ``{SpEx+: A Complete Time Domain Speaker Extraction Network},''
\newblock in {\em INTERSPEECH}, 2020.

\bibitem{spex++}
Meng Ge, Chenglin Xu, Longbiao Wang, Chng~Eng Siong, Jianwu Dang, and Haizhou
  Li,
\newblock ``Multi-stage speaker extraction with utterance and frame-level
  reference signals,''
\newblock {\em IEEE ICASSP}, 2021.

\bibitem{VF_lite}
Quan Wang, Ignacio~Lopez Moreno, Mert Saglam, Kevin Wilson, Alan Chiao, Renjie
  Liu, Yanzhang He, Wei Li, Jason Pelecanos, Marily Nika, and Alexander
  Gruenstein,
\newblock ``{VoiceFilter-Lite: Streaming Targeted Voice Separation for
  On-Device Speech Recognition},''
\newblock in {\em INTERSPEECH}, 2020.

\bibitem{pers_percepnet}
Ritwik Giri, Shrikant Venkataramani, Jean-Marc Valin, Umut Isik, and Arvindh
  Krishnaswamy,
\newblock ``Personalized percepnet: Real-time, low-complexity target voice
  separation and enhancement,''
\newblock in {\em INTERSPEECH}, 2021.

\bibitem{tea-pse}
Yukai Ju, Wei Rao, Xiaopeng Yan, Yihui Fu, Shubo Lv, Luyao Cheng, Yannan Wang,
  Lei Xie, and Shidong Shang,
\newblock ``Tea-pse: Tencent-ethereal-audio-lab personalized speech enhancement
  system for icassp 2022 dns challenge,''
\newblock in {\em IEEE ICASSP}, 2022.

\bibitem{multistage_pse_kuaishou}
Lianwu Chen, Chenglin Xu, Xu~Zhang, Xinlei Ren, Xiguang Zheng, Chen Zhang,
  Liang Guo, and Bing Yu,
\newblock ``Multi-stage and multi-loss training for fullband non-personalized
  and personalized speech enhancement,''
\newblock in {\em IEEE ICASSP}, 2022.

\bibitem{dns_4}
Harishchandra Dubey, Vishak Gopal, Ross Cutler, Ashkan Aazami, Sergiy
  Matusevych, Sebastian Braun, Sefik~Emre Eskimez, Manthan Thakker, Takuya
  Yoshioka, Hannes Gamper, and Robert Aichner,
\newblock ``Icassp 2022 deep noise suppression challenge,''
\newblock in {\em IEEE ICASSP}, 2022.

\bibitem{ecapa_tdnn}
Brecht Desplanques, Jenthe Thienpondt, and Kris Demuynck,
\newblock ``{ECAPA-TDNN: Emphasized Channel Attention, propagation and
  aggregation in TDNN based speaker verification},''
\newblock in {\em INTERSPEECH}, 2020.

\bibitem{voxceleb}
Arsha Nagrani, Joon~Son Chung, and Andrew Zisserman,
\newblock ``Voxceleb: A large-scale speaker identification dataset,''
\newblock in {\em INTERSPEECH}, 2017.

\bibitem{voxceleb2}
Joon~Son Chung, Arsha Nagrani, and Andrew Zisserman,
\newblock ``Voxceleb2: Deep speaker recognition,''
\newblock in {\em INTERSPEECH}, 2018.

\bibitem{voxsrc_19}
Joon~Son Chung, Arsha Nagrani, Ernesto Coto, Weidi Xie, Mitchell McLaren,
  Douglas~A. Reynolds, and Andrew Zisserman,
\newblock ``Voxsrc 2019: The first voxceleb speaker recognition challenge,''
\newblock {\em ArXiv}, 2019.

\bibitem{arcface_cv}
Jiankang Deng, J.~Guo, and Stefanos Zafeiriou,
\newblock ``Arcface: Additive angular margin loss for deep face recognition,''
\newblock {\em IEEE/CVF CVPR}, pp. 4685--4694, 2019.

\bibitem{aam_softmax}
Xu~Xiang, Shuai Wang, Houjun Huang, Yanmin Qian, and Kai Yu,
\newblock ``Margin matters: Towards more discriminative deep neural network
  embeddings for speaker recognition,''
\newblock {\em APSIPA ASC}, 2019.

\bibitem{librispeech}
Vassil Panayotov, Guoguo Chen, Daniel Povey, and Sanjeev Khudanpur,
\newblock ``Librispeech: An asr corpus based on public domain audio books,''
\newblock in {\em IEEE ICASSP}, 2015, pp. 5206--5210.

\bibitem{dnsmos_p835}
Chandan K~A Reddy, Vishak Gopal, and Ross Cutler,
\newblock ``Dnsmos p.835: A non-intrusive perceptual objective speech quality
  metric to evaluate noise suppressors,''
\newblock in {\em IEEE ICASSP}, 2022.

\end{thebibliography}

\end{document}